\documentclass[letterpaper]{ptephy}%%%%%% to generate preprint number with ptep logo

\begin{document}

\newcommand{\gmtwo}{\ensuremath{g\!-\!2}}
\newcommand{\gmtwomu}{\ensuremath{(g\!-\!2)_{\mu}}}
\newcommand{\amu}{\ensuremath{a_{\mu}}}

\newcommand{\TOHOKU}{1}
\newcommand{\TOKYOMET}{2}
\newcommand{\UVIC}{3}
\newcommand{\CHIBA}{4}
\newcommand{\KU}{5}
\newcommand{\TOKYOH}{6}
\newcommand{\TOKYOK}{7}
\newcommand{\KEK}{8}
\newcommand{\RIKEN}{9}
\newcommand{\TRIUMF}{10}

%\linenumbers

%\title{ Enhancement of muonium emission rate from silica aerogel with drilled holes}
\title{ Enhancement of muonium emission rate from silica aerogel
  with a laser ablated surface}

\author{
% \name{P.~Bakule}{\RIKENRAL},
 \name{G.A.~Beer}{\UVIC},
% \name{D.~Contreras}{\TRIUMF},
% \name{M.~Esashi}{\TOHOKU},
 \name{Y.~Fujiwara}{\RIKEN,\TOKYOH},
% \name{Y.~Fukao}{\KEK},
 \name{S.~Hirota}{\KEK,\TOKYOH},
% \name{H.~Iinuma}{\KEK},
 \name{K.~Ishida}{\RIKEN},
 \name{M.~Iwasaki}{\RIKEN},
% \name{T.~Kakurai}{\RIKEN,\TOKYOH},
 \name{S.~Kanda}{\RIKEN,\TOKYOH},
 \name{H.~Kawai}{\CHIBA},
 \name{N.~Kawamura}{\KEK},
 \name{R.~Kitamura}{\RIKEN,\TOKYOH},
 \name{S.~Lee}{\KU},
 \name{W.~Lee}{\KU},
 \name{G.M.~Marshall}{\TRIUMF},
% \name{H.~Masuda}{\TOKYOMET},
% \name{Y.~Matsuda}{\TOKYOK},
 \name{T.~Mibe}{\KEK},
 \name{Y.~Miyake}{\KEK},
 \name{S.~Okada}{\RIKEN},
 \name{K.~Olchanski}{\TRIUMF},
 \name{A.~Olin}{\TRIUMF,\UVIC},
 \name{Y.~Oishi}{\RIKEN},
 \name{H.~Onishi}{\RIKEN},
 \name{M.~Otani}{\KEK},
 \name{N.~Saito}{\KEK,\TOKYOH},
 \name{K.~Shimomura}{\KEK},
 \name{P.~Strasser}{\KEK},
 \name{M.~Tabata}{\CHIBA},
 \name{D.~Tomono}{\RIKEN,
 \thanks{Present Address: Dept. of Physics, Kyoto University, Kyoto, Japan}},
 \name{K.~Ueno}{\KEK},
 \name{K.~Yokoyama}{\RIKEN,
 \thanks{Present Address: School of Physics and Astronomy, Queen Mary University of London, Mile End
Road, London E1 4NS, UK}},
 \name{E.~Won}{\KU},
% \name{S.~Yoshida}{\TOHOKU}
}

\address{
 \affil{\TOHOKU}{Advanced Institute for Materials Research, Tohoku University, Sendai 980-8578,Japan}
 \affil{\TOKYOMET}{Division of Applied Chemistry, Tokyo Metropolitan University, Tokyo, 192-0397, Japan}
 \affil{\UVIC}{Department of Physics and Astronomy, University of Victoria, Victoria BC V8W 3P6, Canada}
 \affil{\CHIBA}{Department of Physics, Chiba University, Chiba 263-8522, Japan}
 \affil{\KU}{Department of Physics, Korea University, Korea}
 \affil{\TOKYOH}{Department of Physics, The University of Tokyo, Tokyo, 113-0033, Japan}
 \affil{\TOKYOK}{Graduate School of Arts and Sciences, The University of Tokyo, Tokyo, 153-8902, Japan}
 \affil{\KEK}{High Energy Accelerator Research Organization (KEK), Ibaraki, 305-0801, Japan}
 \affil{\RIKEN}{RIKEN Nishina Center, RIKEN, Saitama, 351-0198, Japan}
% \affil{\RIKENRAL}{RIKEN-RAL Muon Facility, Rutherford Appleton Laboratory, Harwell Oxford, Didcot, Oxfordshire, OX11 0QX, UK}
 \affil{\TRIUMF}{TRIUMF, Vancouver, BC, V6T 2A3, Canada}
\email{ishida@riken.jp, mibe@post.kek.jp}
}

\begin{abstract}%
  Emission of muonium ($\mu^+e^-$) atoms from a laser-processed
  aerogel surface into vacuum was studied for the first time.
  Laser ablation was used to create hole-like regions with
  diameter of about 270$~\mu$m in a triangular pattern with hole
  separation in the range of 300--500$~\mu$m. 
%revised to reflect referee's comment (Tsutomu Mibe)
%  {\color{blue}
  The emission probability for the laser-processed aerogel sample is at least eight times higher than for a uniform one.
%}
%   The emission
%  probability for the laser-processed aerogel sample can be at
%  least an order of magnitude higher than for a uniform one.
\end{abstract}

\subjectindex{C31, G04}

\maketitle

%\section{Introduction}

The muon \gmtwo\ is one of the fundamental particle properties
for which both theory and experiments can achieve very high
precision.  It has been known there is a discrepancy of about 3.5
$\sigma$ between the best existing measurement from the
Brookhaven experiment (E821) \cite{Bennett2006} and the best
theoretical estimates \cite{HLMNT, Davier}.  We are preparing a
new experiment at J-PARC to measure the \gmtwo\ value based
on a different approach, namely the use of an ultra-cold muon
beam. One of the key steps to the production of the required
intense beam is the ion source, in this case positive muons
($\mu^{+}$) with extremely precise momentum. In
order to accomplish this goal, we have investigated the creation of
neutral muonium atoms ($\mu^{+}e^{-}$, or Mu) with limited
spatial extent in vacuum at room-temperature thermal energies and
with efficiency as high as several \% from a J-PARC low energy muon beam.
The ultra-cold muon beam for the \gmtwo/EDM experiment would result
from laser ionization of Mu in vacuum and subsequent acceleration
of $\mu^{+}$. This paper reports a major step toward our goal
via the use of silica aerogel targets with Mu-emitting surfaces
that were microstructured with a laser ablation technique.

The new approach to be used in the \gmtwo/EDM experiment at J-PARC 
does not need a focusing electric field and
thus removes the constraint of ``magic'' muon momentum (3.094
GeV/$c$) as used in the E821 measurement, so one can use a compact
muon storage ring for 300 MeV/$c$ muons with a high precision
magnetic field based on magnetic imaging technology, and a
compact detection system with particle tracking capability
\cite{E34CDR}. A statistically competitive measurement requires
the production of an ultra-cold muon beam with intensity of the
order of 10$^6$ s$^{-1}$. Positive muons of several MeV energy
and intensity of order 10$^8$ s$^{-1}$ are first stopped
or thermalized near the surface of a suitable material. Some of
the muons are emitted into the vacuum as thermal muonium with
small momentum and energy spread. Laser ionization and
acceleration then create an ultra-cold muon beam.

Utilizing a stopping target with high Mu emission rate is
essential; we have undertaken systematic investigations of
different types and forms of materials.  Silica powder has been
known to be one of the best materials \cite{Beer1986, Woodle1988},
but a major problem with the silica powder is its difficulty in
handling, especially in a high vacuum accelerator environment.
In addition, lack of stability of the bulk powder shape and the
surface condition has been observed to cause emission variations
in long term measurements such as those required for muon \gmtwo/EDM.
We focus our study on silica aerogel, which has a similar
microscopic structure to silica powder, but is rigid and can be
conveniently placed at any orientation in vacuum.

We reported our first measurement of muonium emission from silica
aerogel \cite{Bakule2013} performed at the TRIUMF M15 beam line in
2010 and 2011.
%%% GMM Ishida-san's suggestion, with slight modification
%{\color{blue}
In the measurement, we successfully observed muonium in vacuum from silica
aerogel samples of various densities with yields about 0.3\% per
stopping muon, for emission followed by decay in a region from 10 to 40~mm from
the aerogel surface. This corresponds to a total Mu emission
probability of 1.0\% for decay in vacuum
in a diffusion model interpretation.
%}
%In the measurement, we successfully observed
%muonium emission from silica aerogel samples of various densities
%with yields corresponding to about 1.0\% total emission
%probability in a diffusion model interpretation.
%%%
However, the Mu emission
probability was smaller than published values for silica powders
\cite{Beer1986, Woodle1988} and hot tungsten \cite{Mills1986};
an order-of-magnitude improvement is necessary for
silica aerogel to be considered as a viable alternative source for
an ultra-cold muon beam.

One main conclusion of Ref. \cite{Bakule2013} was that the limitation
to Mu emission was due to the small scale of Mu diffusion
distances compared with the extent of the muon stopping and Mu
formation distribution in aerogel.  The typical distance between
the point of Mu formation to the position of decay of the
$\mu^{+}$ in its 2.2~$\mu$s mean lifetime was at best only about
30~$\mu$m in the aerogel material. This is small compared to the
extent of the muon stopping distribution of about 2~mm, meaning
that only those Mu originating near the flat aerogel surface
could escape from it with significant probability; most muonium
atoms decayed within the aerogel. We considered methods by which
more Mu could be inside the aerogel material at locations closer
to its surface.
\footnote{We refer here to the aerogel surface as that determined
  by the shape of the sample of aerogel. Note that the area of
  the silica surface in aerogel is typically extremely large, up
  to 1000~m$^{2}$~g$^{-1}$ in some of our samples.}
Simulations based on a diffusion model showed that the emission
rate could be increased substantially by an irregular surface,
covered by holes or channels with dimensions of order 100~$\mu$m.

%\section{Aerogel samples}

The highly uniform silica aerogel was produced by the same
methods described in the previous paper
\cite{Bakule2013,Tabata2012}.  We considered and tested several
methods to create a non-uniform structure on this uniform aerogel surface.
Because the silica aerogel is rather fragile, mechanical
treatment of the surface is not considered very promising.  It is
known that laser light can be used for processing
of aerogel material~\cite{Sun2001}.  The laser material processing method for
our targets was developed and tested in RIKEN before the actual
processing was done commercially. % LIGHTEC Inc.

\begin{figure}[t]
\begin{center}
\includegraphics[width=0.7\textwidth, bb= 0 0 403 232]{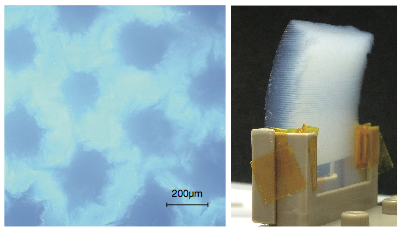}
\caption{ \label{F:aerogel_optical_image} \label{F:aerogel_photo}
(left) An optical image of an aerogel sample
having a triangular hole pattern with equal spacing of 400~$\mu$m.
(right)
 Photograph of a laser-ablated aerogel sample
 installed on a target holder. The muon beam enters the convex
 surface while
 muonium is emitted from the ablated (concave) surface. 
}
\end{center}
\end{figure}

The indentations or holes were created by ablation with a
femtosecond laser with wavelength 800 nm, pulse duration 230 fs,
and energy 0.6 mJ/pulse, at a 1 kHz repetition rate.  The laser
pulse was focused on the surface of silica aerogel with a spot
size of 30 $\mu$m.  The processing time for each hole was 0.8~s
corresponding to 800 pulses.  An aerogel surface area of 30~$\times$~30~mm$^2$
was covered by a triangular pattern of holes with equal spacing
of 300, 400, or 500~$\mu$m.  After ablation, the
hole depths were typically 4.5 to 5.0~mm, compared to the total
thickness of 7~mm for the 29~mg~cm$^{-3}$ aerogel samples.  The
diameter of the holes at the surface of the aerogel was about 270
$\mu$m, enlarged compared to the laser size by the induced
plasma. Surprisingly, the difference in aerogel sample mass
measured before and after the ablation procedure was only 5--10\% of the
expected difference based on hole geometry and aerogel density,
perhaps due to densification of aerogel in the vicinity of the
holes.
Detailed microscopy showed that the holes were not smooth and
regular in shape, but were surrounded by microfractures. 
%and were tapered along their depth.  
Figure \ref{F:aerogel_optical_image}
shows an optical microscope image of the hole pattern, as well as
a photo of one of the targets in a holder prior to exposure to
muons. Note the bend created in the aerogel sample as a result of
contraction after ablation from one side. We expect to be able to
reduce or eliminate this curvature by modification of the laser
process.

%\section{Measurement and analysis}

The identification of muonium emission from aerogel samples with
our apparatus has been described in detail elsewhere
\cite{Bakule2013}, and is summarized here. The position of a
positive muon decaying near the planar target surface can be inferred
via an extrapolation of points on the track of the positron
emitted in the decay. The positron energy should also be measured in
order to select higher energy particles whose multiple scattering
is less severe, thus improving the extrapolation precision. The
time of decay, along with the muon arrival time
in the target, determines a time interval that is the muon lifetime. When
muonium is emitted from the surface of the aerogel sample, this
interval includes the time of flight from the target to the point
of decay. The time interval and position allow us to infer the
component of velocity perpendicular to the surface following emission.

\begin{figure}[t]
\begin{center}
\includegraphics[width=0.9\textwidth]
{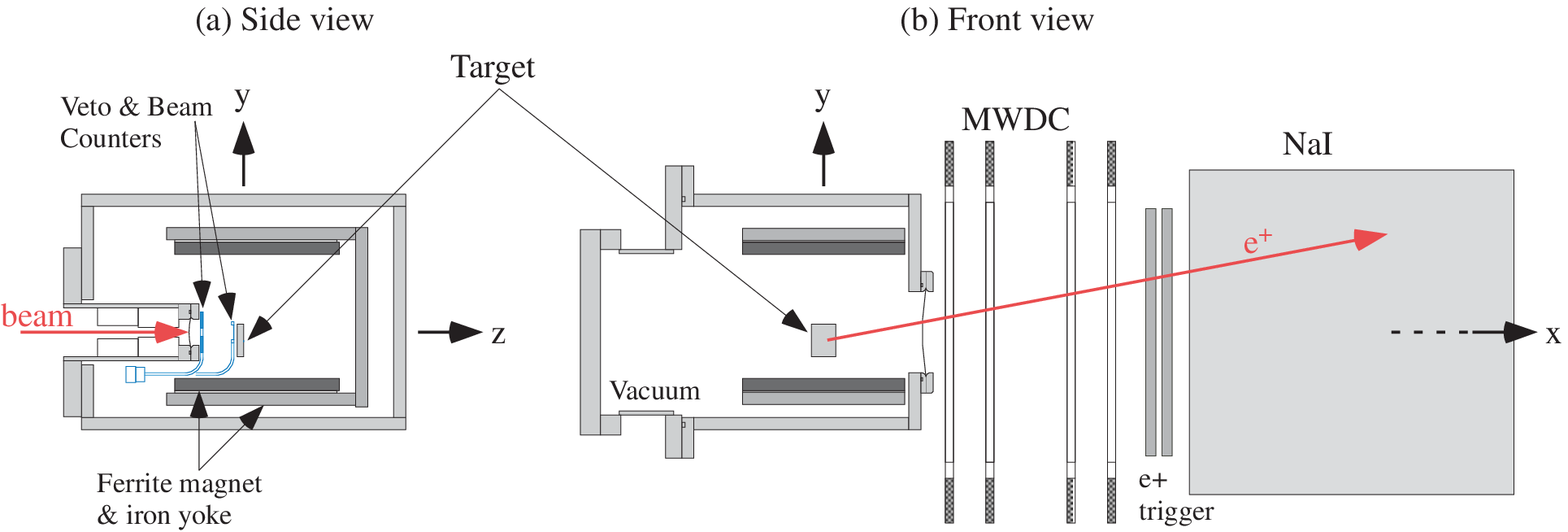}
\caption{\label{F:setup}
Setup for the muonium imaging measurement at the TRIUMF M15 beamline.
The axes of the coordinate system ($x, y, z$) follow right-hand coordinate system as indicated in the figure.
}
\end{center}
\end{figure}

The apparatus, shown in Fig.~\ref{F:setup}, provides an image 
of the muon stopping position and the decay time. Coordinates of the
positron track were determined in the vertical ($y$) and beam
($z$) direction by multiwire drift chambers (MWDCs). A pair of
plastic scintillators ($e^{+}$ trigger) determined the decay time
and the NaI crystal measured the positron energy. The beam of
subsurface muons from the TRIUMF M15 low energy muon beam line
entered the target vacuum system in the $+z$ direction through
collimators and a vacuum isolation window. Muons passed through a
300~$\mu$m beam scintillator into the target sample, whose exit
surface was reproducibly located at $z=0$; there was some
ambiguity introduced by the curvature shown in
Fig.~\ref{F:aerogel_photo}, but not large enough to influence the
results presented here.  The beam intensity was of order
$10^{4}$~s$^{-1}$. Thicknesses of the isolation window and the
scintillator were minimized so that the lowest practical momentum
could be achieved, as this reduces the extent of the muon
stopping distribution for a given beam line momentum resolution
$\Delta p/p$, typically 5\% (FWHM). The targets were
selected to have approximately equivalent mass per unit area to
reduce differences in beam momentum requirements. As a result,
the central momentum was typically near 23.0 MeV/$c$ to stop half
of the beam in a target layer, with only small variations among
targets.

The probability that a muonium atom reaches the surface of the
aerogel target layer and is emitted into vacuum depends on the
location at which the muonium is formed inside the layer; the
nearer the muonium atom is to the surface at the beginning of its
motion in the silica aerogel, the more likely it is to be emitted
from the surface. Simple planar diffusion models of the motion predict
that this probability falls exponentially with initial distance
from the surface for a decaying particle. The exponential has a
mean distance or planar diffusion length $l_{d}=(D\tau)^{-\frac{1}{2}}$,
where $D$ is the diffusion parameter and $\tau$ is the mean
lifetime. In practice, the diffusion length is small compared to
both the thickness of the target and the range spread of muons
stopping in the target. Thus it is expected that the highest rate
of muonium emission into vacuum would be achieved when the
stopping density is maximized within the small distance $l_{d}$ of the
surface. This optimization can be accomplished with the muon
decay position imaging system; it is used to record the relative
rate of muon decays in the target as the mean beam momentum --- thus
the muon stopping position distribution --- is adjusted.

\begin{figure}[t]
\begin{center}
\includegraphics[width=0.9\textwidth]
{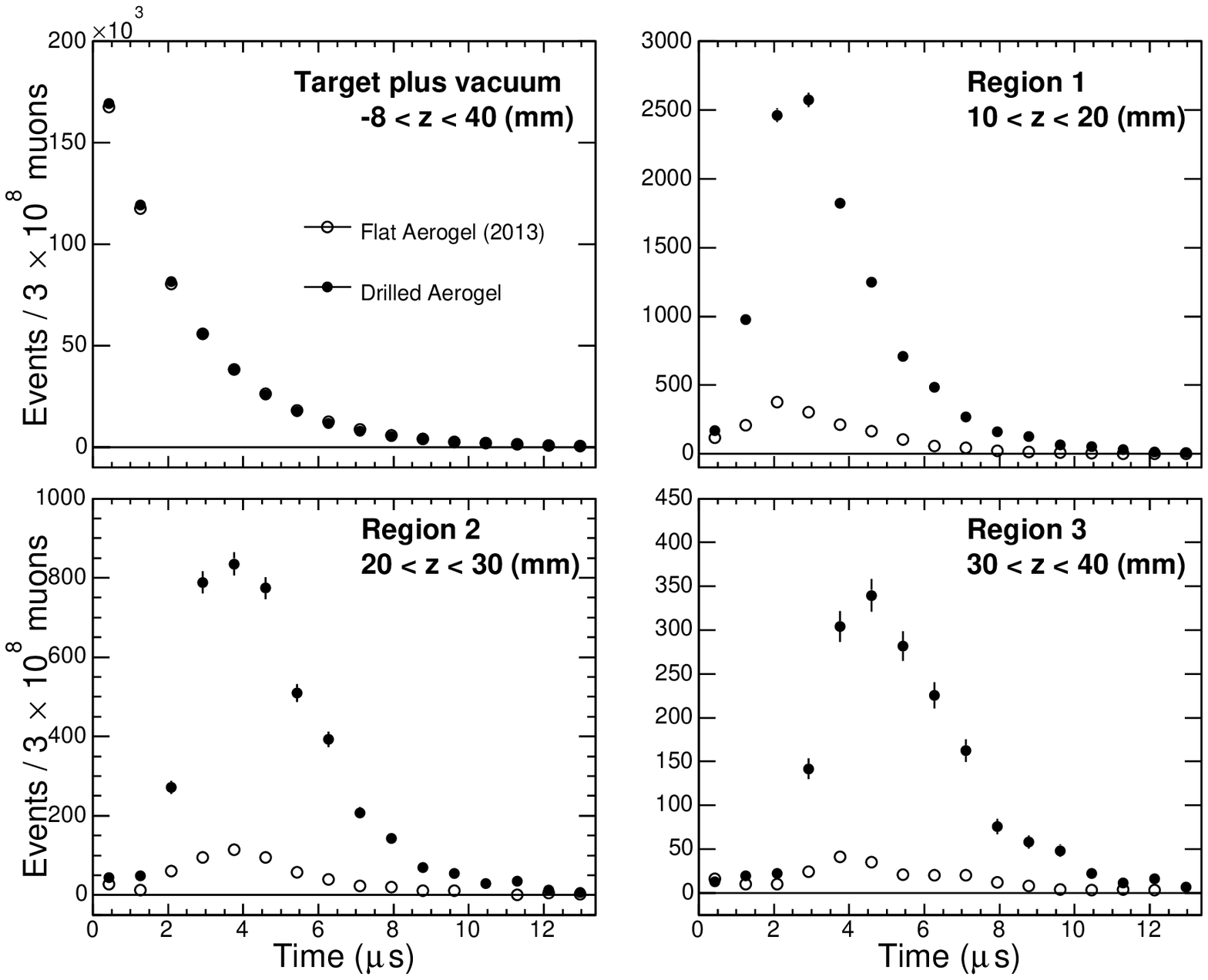}
\caption{\label{F:tdist}
Time distributions of positrons in the entire target region and
in each of three vacuum regions, for flat aerogel (open circles) and laser-ablated aerogel 
with pitch of 300~$\mu$m (closed circles). No background has been
subtracted.
}
\end{center}
\end{figure}

\begin{table}[t]
\begin{center}
\caption{
\label{T:yield}
Yield of Mu in the vacuum region 1--3. For all laser processed
samples, the diameter of the structure is 270~$\mu$m.}
\begin{tabular}{ l c r}
\hline
Sample &  Laser-ablated structure & Vacuum yield\\
& (pitch) & (per $10^3$ muon stops)\\
\hline
Flat  & none & $3.72 \pm 0.11$\\
Flat (Ref.~\cite{Bakule2013}) & none & $2.74 \pm 0.11$\\
%Flat (high transparency) & none & $3.62 \pm 0.11$\\
%Flat (non-hydrobobic) & none & $2.86 \pm 0.12$\\
%Push-pin drilled & 0.5~mm, 1~mm & $5.56 \pm 0.17$\\
Laser ablated  & 500~$\mu$m & $16.0$~$\pm$~0.2~~\\
Laser ablated  & 400~$\mu$m & $20.9$~$\pm$~0.7~~\\
Laser ablated  & 300~$\mu$m & $30.5$~$\pm$~0.3~~\\
%Laser ablated  & 500~$\mu$m & $15.98 \pm 0.23$\\
%Laser ablated  & 400~$\mu$m & $20.85 \pm 0.73$\\
%Laser ablated  & 300~$\mu$m & $30.50 \pm 0.32$\\
\hline
\end{tabular}
\end{center}
\end{table}

The position resolution of the positron track extrapolation was
estimated as $\sim$2~mm (RMS) using data taken with a target
designed specifically for calibration
and background estimation, a silica plate of thickness
100~$\mu$m.  The time distribution of the positron tracks was
analyzed in four $z$ regions.  The first region, defined to
include decays both from the target and up to 40~mm into the vacuum following the
target, included the entire range ($-8 < z < 40$~mm) considered in
this analysis.
Vacuum regions 1, 2, and 3 were defined as 10~mm wide ranges of
$z$ starting from 10, 20, and 30~mm respectively from the emitting surface of
the target ($z = 0$). The time distributions for the flat
aerogel and ablated aerogel within these regions are shown in
Fig.~\ref{F:tdist}.  The time distribution appears mostly
exponential for decays of muons or Mu from the entire region.
The Mu in vacuum\footnote{Note that the interpretation of the
  vacuum decay events as arising from non-neutral forms
  ($\mu^{+}$) is excluded; a vertical magnetic field of 8 mT was
  present in all measurements that would cause thermal charged
  forms to curl back to the target surface via cyclotron
  motion.}, on the other hand, moves across regions 1--3 with a
thermal velocity.  The time distribution of such Mu is a
convolution of the emission time for Mu to escape the aerogel
sample and the flight time determined by the velocity
distribution, creating the peak structure in the regions 1--3.
There are small contributions in regions 1--3 from muon decay
events in the target that were subtracted by assuming the
exponential functional form in order to estimate the yield of
muonium in vacuum.

Table~\ref{T:yield} summarizes the Mu yield, after subtraction of
the background, summed for regions 1--3.  The beam momentum was
set to stop about 50\% of muons in the sample; the remainder
mostly escaped from the target and vacuum regions where their
decays were not detected.  
%%% GMM Added in response to referee's point 7. It should precede
%%% the three sentences added by TM. I estimate 4% based on the
%%% laser30_300 comparison of target position (slide 9 in July 8
%%% presentation) which showed a shift of 0.4 mm, combined with
%%% the estimate of a bias of -10% per mm shift as shown in the
%%% slides I sent by email regarding this systematic. This bias
%%% could be corrected in order to reduce the systematic
%%% uncertainty, but we haven't done that in this analysis and I
%%% do not propose to change it now.
%{\color{blue}
The quoted uncertainties are statistical only. Most systematic
uncertainties are removed by the model-independent yield
analysis; those that remain are estimated to
be at the level of 4\% from the laser-ablated targets, dominated
by the effect of curvature of the emitting surface, and about 1\%
for the flat samples.
%}
% added to reflect referee's comment (Tsutomu Mibe)
%{\color{blue}
The Mu yield from the flat sample is about 40\% higher
than that in Ref.~\cite{Bakule2013}. The densities and
geometrical dimensions are nearly identical for both
samples. However, they were manufactured by slightly
different processes; supercritical ethanol drying at
260~$^{\circ}$C was used in Ref.~\cite{Bakule2013}, while
supercritical carbon dioxide drying at 80~$^{\circ}$C
was used for the samples in this work. This could lead to the difference in the Mu yield for the slightly different
flat samples.
%}
An enhancement of Mu in vacuum from the laser ablated aerogel
compared to flat aerogel is evident.  The yield is higher when
the hole pitch is smaller. The highest yield observed among these
targets was the laser ablated sample with 270~$\mu$m diameter and
300~$\mu$m pitch. That yield is 3\% compared to the total number
of muons observed to decay in the combined target and vacuum regions.

%\section{Discussion and prospects}
The application of this result to development of a muonium
production target in the \gmtwo/EDM experiment at J-PARC is
discussed in the following.  The beam momentum and its spread at
J-PARC is designed to be 28~MeV/$c$ and 5\% (RMS), respectively.
The projected yield of muonium at J-PARC is estimated as 0.01
per incident muon under the assumption that only a small region
near the surface contributes to emission~\cite{Bakule2013}.
Taking into account the area of overlap of muonium in vacuum with
the ionizing laser, and the ionization efficiency~\cite{E34CDR},
the estimated ultra-slow muon rate is $0.2\times 10^{6}$/s.  
% added to reflect referee's comment  by Tsutomu Mibe
%{\color{blue}
This is five times smaller than the design intensity to achieve the final statistical 
sensitivity of 0.1~ppm on \gmtwo. Further improvement on the muonium yield is necessary
to reach the final sensitivity.
Nevertheless, the projected statistical sensitivity with the current muon source is 0.4~ppm
for \gmtwo\ in $10^7$~s of data taking time with 50\% beam
polarization, exceeding the precision of the previous
measurement~\cite{Bennett2006}.
%}

%\section{Conclusions}
A technology was developed to introduce non-uniform structure
(holes) on the surface of silica aerogel samples with laser
ablation.  Emission of muonium into vacuum increased in all
laser-treated samples tested.  The emission rate for the ablated
aerogel with holes of pitch 300~$\mu$m is
%revised to reflect referee's comment by Tsutomu Mibe
%{\color{blue}
at least eight times higher than the one without the laser treatment.
%}
%an order of magnitude higher than the one without the laser treatment.

%\section{Acknowledgment}
%\section*{Acknowledgment}

 The authors are pleased to acknowledge the support from TRIUMF
 to provide a stable beam during the experiment.
 Special thanks go to R. Henderson, R. Openshaw, G. Sheffer, and M. Goyette
 from the TRIUMF Detector Facility.
 We also thank D. Arseneau, G. Morris, B. Hitti, R. Abasalti, and D. Vyas 
 of the TRIUMF Materials and Molecular Science Facility.
 Practical advice and laser equipment given by Yuichi Asakawa of LIGHTEC Inc. 
 has been a great help in fabrication of laser ablated aerogel.
 Research was supported in part by the MEXT KAKENHI Grant Number 2318005 (Japan), 
 NSERC Discovery Grant (Canada), and Center for Korean J-PARC Users No. NRF-2013K1A3A7A06056592 (Korea).


%\bibliographystyle{ptephy}
%\bibliography{sample}
%
% once the .bbl file has been generated then place the text in your article.

%\vfill\pagebreak

\end{document}